Review Paper

# Challenges in Developing Secure Mobile Health Applications - A Systematic Review

Bakheet Aljedaani [1,2], M. Ali Babar[1,3]

[1]Centre for Research on Engineering Software Technologies, School of Computer Science, The University of Adelaide, Australia
[2]Computer Science Department, Aljumum University College, Umm Alqura University, Makkah, Saudi Arabia
[3]Cyber Security Cooperative Research Centre (CSCRC), Australia

Corresponding Author:
Bakheet Aljedaani
School of Computer Science
The University of Adelaide
Adelaide SA 5005
Australia
Phone: +61 8 8313 4729
Email: bakheet.aljedaani@adelaide.edu.au

## Abstract

**Background: Mobile health (mHealth) applications (apps) have gained significant popularity over the last few years due to its tremendous benefits, such as lowering healthcare cost and increasing patient awareness. However, the sensitivity of healthcare data makes the security of mHealth apps a serious concern. Poor security practices and lack of security knowledge on developers' side can embed several vulnerabilities in mHealth apps.**

**Objective: In this review paper, we aim to identify and analyse the reported challenges that the developers of mHealth apps face concerning security. Additionally, our study aimed to develop a conceptual framework with the challenges faced by mHealth apps development organization for developing secure apps. The knowledge of such challenges can help to reduce the risk of developing insecure mHealth apps.**

**Method: We followed the Systematic Literature Review (SLR) method for this review. We selected studies that have been published between January 2008 and October 2020 since the major app stores launched in 2008. We selected 32 primary studies using predefined criteria and used thematic analysis method for analysing the extracted data.**

**Results: Out of 32 studies, we identified nine challenges that can affect the development of secure mHealth apps. These challenges are 1) lack of security guidelines and regulations for developing secure mHealth apps, 2) developers' lack of knowledge and expertise for secure mHealth app development, 3) lack of stakeholders' involvement during mHealth app development, 4) no/little developers' attention towards the security of mHealth app, 5) lack of resources for developing secure mHealth app, 6) project constraints during mHealth app development process, 7) lack of security testing during mHealth app development, 8) developers' lack of motivations and ethical considerations, and 9) lack of security experts' engagement during mHealth app development. Based on our analysis, we have presented a conceptual framework which highlights the correlation between the identified challenges.**

**Conclusion: While mHealth apps development organizations might overlook security, we conclude that our findings can help them identify the weaknesses and improve their security practices. Similarly, mHealth apps developers can identify the challenges they face to develop mHealth apps that do not pose security risks for users. Our review is a step towards providing insights into the development of secure mHealth apps. Our proposed conceptual framework can act as a practice guideline for practitioners to enhance secure mHealth apps development.**

## KEYWORDS

## Introduction

### Background

The use of mobile applications (apps) in healthcare has gained widespread adoption [1, 2]. Lack of health professionals, especially in a rural area, is an excellent motivator for mobile health (mHealth) apps adoption [3]. mHealth apps rely on portability and context-sensitivity of mobile computing to improve access to healthcare services that are cost-effective, scalable, and pervasive [4]. Leveraging mHealth apps would improve access to healthcare services, lower its cost and increase patients' health awareness [5]. According to the World Health Organization (WHO), mHealth is defined as *"medical and public health practice* supported by mobile devices, such as mobile phones, patient monitoring devices, personal digital assistants (PDAs), and other wireless devices [6]". There are several types of mHealth apps developed for health purposes ranging from general health apps such as decision, support, vitals, and reproductive health apps; through fitness apps for an activity tracker, nutrition tracker and mindfulness [5]. The number of mHealth apps has grown massively after launching the centralized mobile apps repositories (i.e., Google Play and Apple Store) in 2008. It has become easier for mobile developers to distribute their apps to a wide range of users [7]. Research2guidance, an organisation for providing research and consultancy for digital health, reports that 78,000 new mHealth apps were added to the app stores in 2017. The report also showed that mHealth apps downloads reached 3.7 billion and the market revenue for digital health reached USD 5.4 billion in 2017 [8].

Security of mobile apps generally and mobile health apps, in particular, becomes one of the primary concerns since mobile apps are more vulnerable to attacks [9]. Most mobile apps can collect, process, store, and transmit user and device data in and out of a device over various networks [5]. Compromising the confidentiality, integrity, and availability of such data would lead to severe consequences including but not limited to compromised devices data, and leading to financial loss [10]. In mHealth apps, the security becomes a significant concern due to health-critical data privacy and integrity [5, 11]. An attack to falsify clinical measurements can lead to unnecessary care for patients as they think they are sicker than they actually are, to cause medical, legal, and social consequences [12].

Health professionals are increasingly relying upon health data which are collected via mHealth apps to make their decisions, such as dermatologic care [13], chronic management [14, 15], and clinical practices [16]. Data manipulation can significantly impact the treatment causing serious results, e.g., worsened morbidity or death [17, 18]. Whilst health regulations and laws (i.e., The US Health Insurance Portability and Accountability Act - HIPAA, European General Data Protection Regulation - EU GDPR) strive to protect medical integrity and patients' privacy by focusing on hospitals, doctors and insurance firms. Little attention has been paid to support mHealth apps developers by providing them with suitable guidelines for developing secure apps [5, 19].

A large part of mHealth apps' security relies on developers'[1] experience for designing and developing secure apps. According to [1, 12, 15, 20, 21], most mHealth apps have not fully implemented mechanisms to protect health data. Studies also claimed that mHealth developers might fail to appropriately implement basic security solutions such as authentication, encryption for data at rest and data in transit. It is being recognised that it is critically important to thoroughly train mHealth apps developers in implementing suitable security mechanisms to protect patients' data from being stolen or compromised [12, 20]. Hence, it is crucial to identify and synthesize the reported challenges of developing secure mHealth apps as a body of knowledge for research and practice. We have reviewed the relevant literature to determine the security challenges by focusing on the developers rather than the solutions that can be applied. Our Research Question (RQ) for this literature review is: *What are the challenges that developers of mHealth apps face with respect to implementing security?*

**Previous Work**

The challenges of developing secure software have been receiving increasing attention in recent years. A review by Kanniah et al. [22], which included 44 studies, has identified the factors that influence secure software development practices. The study finds that security skills, expertise, tools and development time are among the factors that impact on secure software development. The identified factors were classified into institutional context, people and action, project content, and software development process factors. Thomas et al. [23] address the issues that security auditors face during the application review for security bugs. The study recommends further support for the development process by providing security-related tools and effective communication tools for developers' interaction. Further support for software developers has also been recommended by providing motivation (e.g., reward or recognition) as well as provide solutions for technical challenges such as using third-party libraries issues. The authors recommended recruiting security experts within teams and make them available for answering questions. Raghavan et al. [24] present a model for achieving security during the Software Development Lifecycle (SDLC). Their model suggests the following factors: security policy, management support, security-related training for developers and development process control. Weir et al. [25] studied the positive factors that enhance the development of secure software. The work identified the interventions that lead to achieving security by performing a threat model, organising motivational workshops to engage team members, a continuous reminder of developers. The study also highlighted other intervention that needs to be considered, such as components choice of security tools, performing static analysis, developers training, and performing penetration testing and code review.

Some studies also aimed to help mobile apps developers to develop secure apps by providing guidelines for the development process [6, 26, 27]. Given the increasing realization of the need to provide mHealth apps' developers with appropriate knowledge/training and support for developing secure apps, there is a critical need to identify and analyse the challenges that prevent mHealth apps' developers from developing secure apps. Our findings would contribute to a body of knowledge about the challenges that

---

[1]The term developer has been used in this paper referring to professionals who are engaged with engineering and development of mHealth apps.

mHealth apps' developers face with respect to security.

**Comparison with Prior Studies**

Prior reviews, e.g. [28, 29], were more focused on investigating the security measures and technical solutions employed by developers. However, a few challenges were raised in [28, 29]. Katusiime et al. [28] systematically reviewed privacy and usability issues and solutions in mHealth systems. The study considered developers' lack of security knowledge, lack of security framework as external factors that need to be considered. Another review by Marquez et al. [29] was more on the security issues of telehealth systems. The study focused on classifying security (i.e., attacks, vulnerabilities, weaknesses, and threats), and presenting security strategies (i.e., detect attacks, stop or mitigate attacks and react to attacks) of telehealth systems. Also, the study reported some security practices that need to be ensured, such as having discussion about architectural styles (e.g., security patterns), and engaging stakeholders during the development of the app. To the best of our knowledge, there is no SLR that explicitly investigate the challenges that faced by mHealth apps developers to implement the security of mHealth apps. Thus, we aim to fill the gap and provide insights into the development of secure mHealth apps.

## *Research Method*

This research has been undertaken as a Systematic Literature Review (SLR). It is one of the most widely used research methods of Evidence-Based Software Engineering (EBSE). SLR provides a well-defined process for identifying, evaluating, and interpreting all available evidence relevant to particular research. We followed the guideline of Barbara Kitchenham to perform an SLR that involves three main phases: defining a review protocol, conducting the review and reporting the review [30]. In this section, we briefly describe the main components of the review protocol and its conduction. Our review protocol has six components including (i) research question (ii) search strategy (iii) data source (iv) study selection process (v) inclusion and exclusion criteria, and (vi) data extraction and data synthesis. Figure 1 shows an explanation for components (i) – (v).

**(i) Research Question**

Our review's objective was to identify and codify the challenges that hinder mHealth apps' developers from developing secure apps, as in Figure 1 (i). This review's findings would enable us to identify the potential gaps that need to be further investigated based on the developers' perspectives.

**(ii) Search Strategy**

We used the following strategies to form our search string: (1) identifying the major terms based on the study focus and the research question, (2) identifying all the possible keywords and related synonyms based

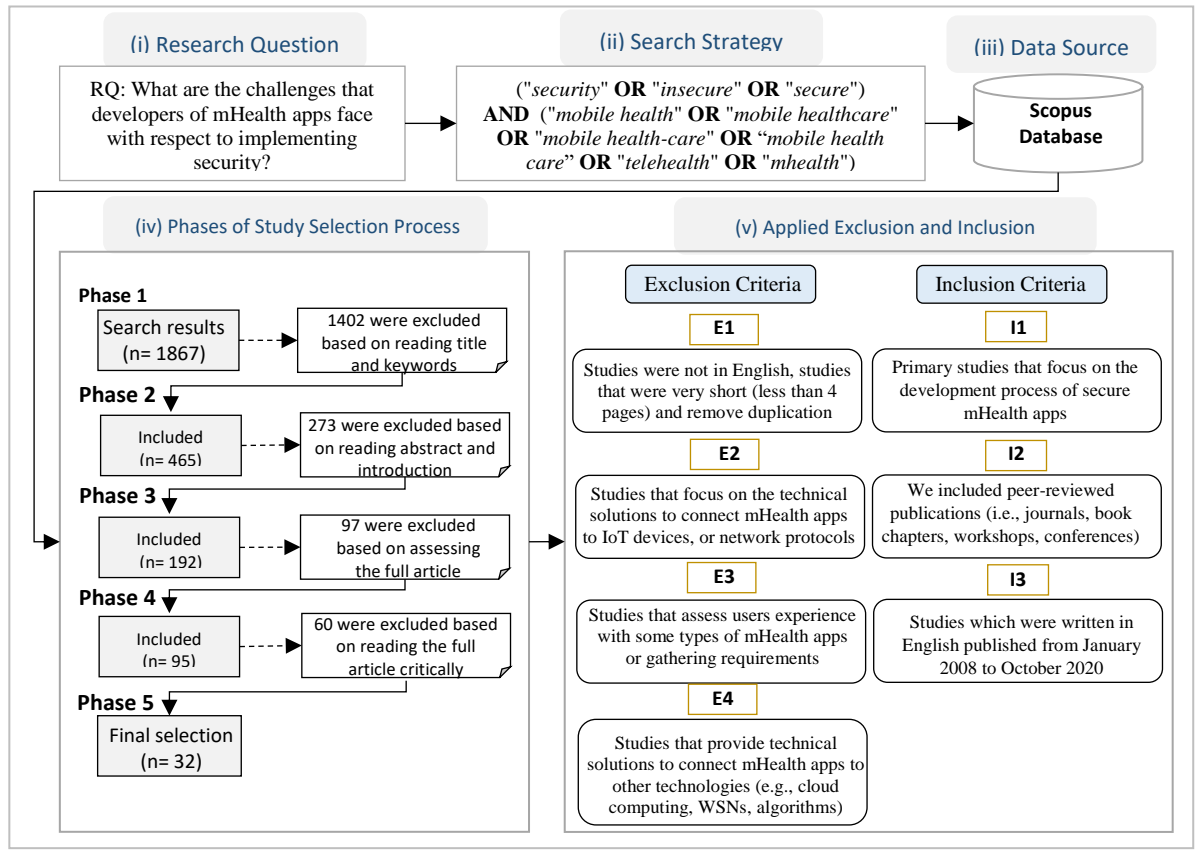

**Figure 1.** Overview of the Research Methodology
*[(i) Research Question, (ii) Search Strategy, (iii) Data Source, (iv) Phases of Study Selection Process, (v) Applied Exclusion and Inclusion Criteria]*

on our experience and previous work, (3) using the Boolean "AND" to join major terms and the Boolean "OR" to join alternative terms and synonyms. Figure 1 (ii) shows the used search string for this review.

### (iii) Data Source

As in Figure 1 (iii), we used Scopus digital library as our primary search library as there are many successful examples of other researchers, e.g., [30], for limiting their search to Scopus. The Scopus indexing system has the advantages of facilitating the formulated complex search string, frequently updated and keeping track of a large number of journals and conferences in software engineering studies. Furthermore, Scopus is an indexing database which provides, name, keywords, and abstract for all published articles. Any pointed articles can be further searched and downloaded to review the whole article regardless of which database it actually exists.

### (iv) Studies Selection Process

As illustrated in Figure 1 (iv), we performed several criteria for excluding studies in our SLR. Further details would be presented as follows:

*Phase 1- Automatic search*: We ran our search string on Scopus digital library. Thus, we retrieved a total of 1867 potential articles.

*Phase 2- Title and keyword-based selection:* We carefully reviewed the title and keywords to decide whether each of the retrieved articles was relevant to our SLR. We retained the papers for the next inspection when we could not decide by reading the titles and keywords. Thus, we excluded 1402 articles and included 465 articles for the next phase.

*Phase 3- Abstract and introduction-based selection:* We looked into the abstract and introduction for each article. This phase enabled us to include 192 articles and discard 273 articles.

*Phase 4- Full paper scanning-based selection:* We scanned the full article to ensure that they are relevant to our SLR objective. Thus, we included 95 articles and excluded 97 articles.

*Phase 5- Critical review based selection:* We critically reviewed the included papers and excluded the duplication (e.g., extended versions of the studies were included, and shorter versions were excluded). Thus, we excluded 60 articles and included 32 studies, referred to as S1 to S32. (Appendix 1).

### (v) Inclusion and Exclusion Criteria

Figure 1 (v) shows our Inclusion (I) and Exclusion (E) criteria that we applied for papers selections. We included:

- Primary studies that focus on the development process of secure mHealth apps.

- We studies which were written in English published from January 2008 to October 2020 since major app stores (Google Play and Apple Store) were launched in 2008.
- We included peer-reviewed publications (i.e., journals, conferences, workshops and book chapters.

Besides excluding non-peer-reviewed studies (i.e., lecture notes, summaries, panels, and posters) and the studies which were not written in English, we excluded studies that contained irrelevant content for our review such as,

- Studies that focus on investigating technical solutions (e.g., encryption methods, authentication mechanisms, access control) for mHealth apps.
- Studies provide technical solutions to connect mHealth apps to Internet of Things (IoT) devices or cloud computing technology.
- Studies that focus on sensors layer (e.g., Wireless Sensor Network - WSNs), developing algorithms, or network protocols for mHealth apps.
- Studies that focus on mHealth apps quality or gathering functional requirements.
- Studies that examine users experience with some mHealth apps (e.g., patient management apps).

### (vi) Data Extraction and Synthesis

We divided the extracted into two types, demographic data, and challenges of developing secure mHealth apps. Our data extraction form is shown in Appendix 2. We performed descriptive statistics to analyze demographic data. For answering our RQ, we used Endnote tool to manage bibliography and utilised Excel spreadsheets to extract and synthesise data. We used thematic analysis, a qualitative analysis technique, to analyse and synthesize the extracted data for deriving the results for this review [31]. We mainly followed the thematic analysis method's five steps as detailed below: (1) Familiarizing with data: In this step, we tried to read and examine the extracted data items. (2) Generating initial codes: In this step, we extracted the initial lists of challenges. (3) Searching for themes: We tried to combine different initial codes generated from the second step into potential themes. (4) Reviewing and refining themes: the identified challenges from step three were checked against each other to understand what themes had to be merged with others or dropped. (5) Defining and naming themes: In this step, we defined a name for each challenge. Figure 2 demonstrates an example, which was taken from (S4), of how our final list of challenges was identified.

To further enhance our analysis, we developed a conceptual framework to present the correlation among the identified challenges. We followed the steps of Patrick Regoniel to develop a conceptual framework that involves four steps: choose the topic, do a literature review, isolate the important variables, and generate the conceptual framework [32]. It should be noted that the initial coding was done by the first author and was reviewed and revised (followed by a discussion wherever required) by the second author to avoid potential bias.

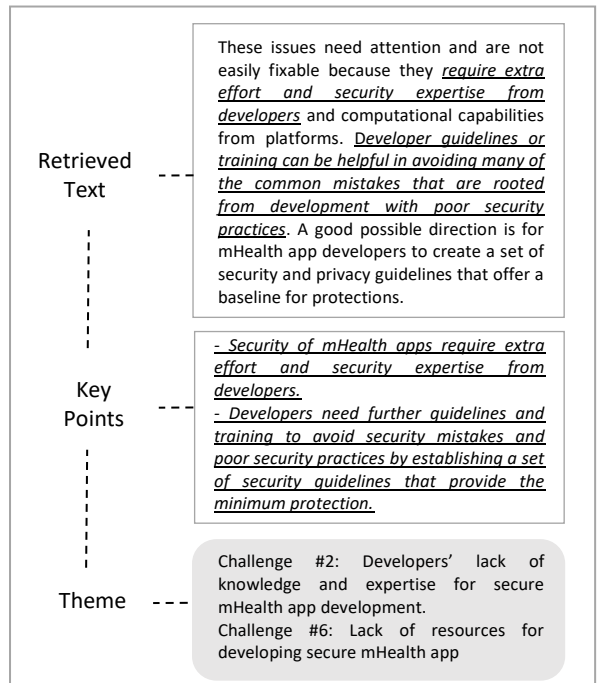

**Figure 2.** Example of the Steps of Applying Thematic Analysis on Qualitative Data

## Results

We now present the findings of our SLR. We classified the findings into (A) demographic information, (B) challenges of developing secure

mHealth apps, and (C) conceptual framework for the identified challenges.

### A. Demographic Information

In this subsection, we present the demographic information based on the venues (i.e., journal, conference, or workshop) of the selected studies, as shown in Figure 3.

Providing such information would be useful for new researchers interested in conducting research on this particular area. We selected 32 primary studies for this review, and the complete list of the reviewed articles is available in Appendix 1. All selected studies were mainly discussing the security aspects of mHealth apps. Figure 3 shows the distribution, year of publication, and the different venues. It should be noted that our reviewed studies were published from 2012 to 2020. Out of 32 studies, we noticed that 23 studies (72%) were published as journal papers. 7 studies (22%) were published in conferences, while 2 studies (6%) were published as workshop papers. Furthermore, we noticed that 11 studies (34%) were published in JMIR (i.e., Journal of Medical Internet Research, and JMIR mHealth and uHealth), and 2 studies were published in the International Conference on Future Internet of Things and Cloud Workshops (2017, 2019).

### B. Challenges of Developing Secure mHealth apps

This subsection reports the results based on our analysis to answer the study RQ, *What are the challenges that developers of mHealth apps face with respect to implementing security?* Our analysis has identified 9 challenges (referred to as C1 to C9) that hinder apps' developers from developing secure mHealth apps. The identified challenges were ordered based on their frequency within the reviewed studies. The findings are detailed below. Table 1 illustrates the identified challenges, the key points which lead us to consider them from the reviewed studies, and challenges frequency.

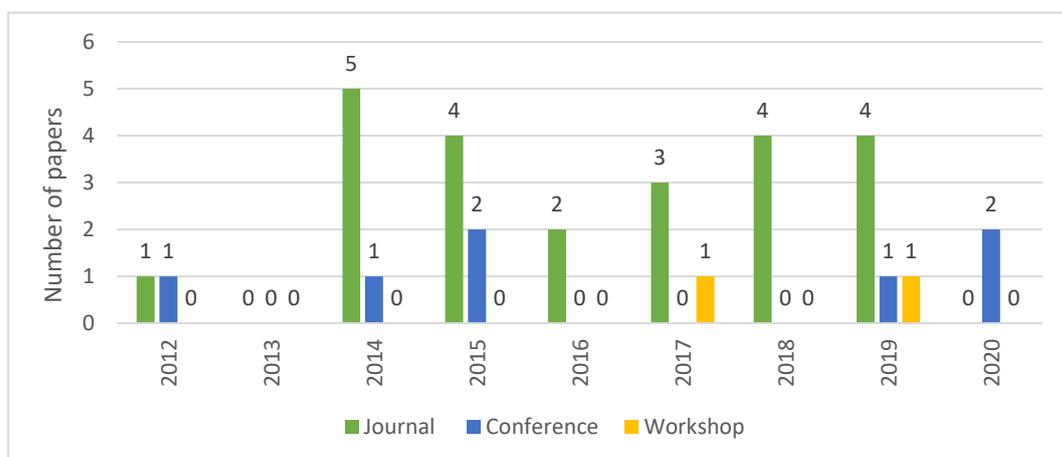

**Figure 3**. The Number of Selected Studies Published per Year and their Distribution over Types of Venues

**Table 1.** Challenges in Developing Secure mHealth Apps

[Frequency, in the table, refers to the number of studies that indicate identified the challenge (n=32)]

| Challenge # | Key Points from Reviewed Studies | Frequency (%) |
|---|---|---|
| C1. Lack of security guidelines and regulations for developing secure mHealth app | ▪ Lack of security guidelines, regulations, direct laws about the security requirements, secure designing, security testing, security features that need to be employed in mHealth apps (S4, S5, S6, S7, S10, S12, S13, S15, S16, S20, S22, S23, S26, S29, S31)<br>▪ Lack of framework or standards (e.g., standardized policies and methodologies to ensure the security standards are met), for developing secure mHealth apps (S2, S3, S29, S31)<br>▪ Lack of compliance with the available guidance and/or standard (S25, S29)<br>▪ Challenges for the developers to deal with legal obligations, policies, and procedures (S32) | 20 (63%) |
| C2. Developers' lack of knowledge and expertise for secure mHealth app development | ▪ Insufficient knowledge of software developers about the security risks of mHealth apps (S12, S17, S18, S27)<br>▪ Lack of developers' security awareness (e.g., towards the potential threats of mHealth apps (S3, S9, S14, S21, S28, S32)<br>▪ Developers lack of knowledge towards secure coding practices, using secure APIs, and utilizing up-to-date libraries (S18) or secure third-party services by mHealth apps developers that could misuse users' health data (S1, S11, S19, S24)<br>▪ Developers lack of knowledge towards utilizing security measure (e.g., (e.g., TLS security for servers, proper protection for user passwords ) of mobile devices (S3, S8, S22, S25)<br>▪ Lack of experience in secure software development for the developers (S4)<br>▪ Lack of auditing security knowledge and review what knowledge they have (S25). | 18 (56%) |
| C3. Lack of stakeholders' involvement during mHealth app development | ▪ Lack of stakeholders' participation during the development lifecycle of mHealth apps (S5, S10, S20, S29, S30)<br>▪ Lack of security understanding by health professionals when they engage in the development process causing poor elicitation of security requirements (S5) | 6 (19%) |
| C4. No/little developers' attention towards the security of mHealth app | ▪ Developer' assumption that users are not concerned about security (S32)<br>▪ Security is not developers concern (S11, S21)<br>▪ Security issues should be resolved by the testers (S32)<br>▪ Developers with no security focus skip all security measures (S18)<br>▪ Developers are not considering secure design principles and privacy guidelines (S31) | 5 (16%) |
| C5. Lack of finanical resources for developing secure mHealth app | ▪ No/low budget assigned for employing security measures (S32)<br>▪ Unavailability of security tools (S32)<br>▪ Developers lack of training about developing secure mHealth apps (S4, S5)<br>▪ Lack of research and development efforts to facilitate developing secure mHealth apps (S14) | 4 (13%) |
| C6. Time constraints during mHealth app development process | ▪ Rushing to the market which leaves vulnerabilities in mHealth apps (S18, S26, S32)<br>▪ The long process of gaining consent or approving the development choices of the developers (S7) | 4 (13%) |
| C7. Lack of security testing during mHealth app development | ▪ Lack of conducting security testing (S32)<br>▪ Lack of conducting proper security testing (e.g., vulnerability scan) for mHealth apps (S6, S18, S23) | 4 (13%) |
| C8. Developers' lack of motivations and ethical considerations | ▪ Lack of motivations for developers during the development process of mHealth apps (S27)<br>▪ Developers lack of ethics during the development process of mHealth apps (S10, S30) | 3 (9%) |
| C9. Lack of security experts' engagement during mHealth app development | ▪ Lack of collaboration and discussion with security experts from the beginning of the development lifecycle of mHealth apps (S18, S32) | 2 (6%) |

### C1: Lack of security guidelines and regulations for developing secure mHealth app

Security guidelines refer to a set of suggested actions or recommendations for things to do or avoid during software development [33]. The security guidelines help apps developers, mostly inexperienced, adopt effective security practices and write secure codes. They contain accessible information, properly layered and searchable, with good coverage of all security aspects (e.g., cryptography, handling user input and privileges [26]). It would be ideal for clarifying that there are numerous security guidelines for ensuring the security of mobile apps (e.g., Open Web Application Security Project – OWASP). According to Nurgalieva et al. [34], the available security guidance for developing secure mHealth apps can be categorized into: 1) guidelines, recommendations, or principles, 2) app development practices (i.e., applied security mechanisms) to ensure mHealth security, 3) models of user behaviour and preferences related to security and/or privacy. Such guidelines (e.g., GDPR) have had the effect of raising awareness and establishing a minimal set of expectations. However, they do not address the issue of the development of systems which meet privacy and security requirements [34]. Additionally, Assal et al. [35] indicated that that security guidelines do not exist or are not mandated by the companies, or that developers might lack the ability or proper expertise to identify vulnerabilities despite having general security knowledge. Our reviewed studies, including (S3, S4, S12, S20) have pointed out a general lack of security guidelines for developing secure mHealth apps. Zubaydi et al. call for effective guidelines that can help developers build secure mHealth apps (S12). Even though there are guidelines to protect health data (i.e., HIPAA guides), they do not provide specific instructions for developing secure mHealth apps. Furthermore, it has also been claimed that there is a lack of security frameworks, standards, compliance checklists and regulations (S22, S18, S13, S20, S2, S9). Legal restrictions (i.e., obtaining security certification) ensure that mHealth apps development organisations are not developing vulnerable mHealth apps (S11, S12).

### C2: Developers' lack of knowledge and expertise for secure mHealth app development

Security knowledge of mobile apps developers plays a significant part in developing secure mHealth apps. Lack of security knowledge would result in creating an insecure app that leaks health-critical data to attackers. The reviewed studies indicate that mHealth app developers do not have enough security education covering important security aspects. Consequently, developers follow insecure programming practices (e.g., employing improper security solutions) (S22, S19, S25), and/or improper handling for mHealth apps permissions (S23). Furthermore, developers' lack of security knowledge leads to make wrong security choices when attaching particular device with mHealth apps (e.g., tracking device that help to monitor user behaviour) (S11, S12, S18), or integrating an app with other systems (S13). Making an incorrect security decision may allow health apps to share health-critical data with other mobile apps, untrusted apps or external host (S12). In fact, mHealth apps developers were found that they make their security decisions based on their best assumption or strategies (S24). Thamilarasu et al. (S18) conducted a vulnerability scan that has reported 248 vulnerabilities in the top 15 Android-based mHealth apps. The study revealed that the top three most common vulnerabilities were not errors in the system, but instead, errors in the developers' choices (i.e., selecting suitable cipher, choice of permissions to request on a mobile device). The study concluded that most vulnerabilities could have been prevented through proper coding and secure engineering practices.

Keeping in mind that threats landscapes are changing rapidly; thus, dealing with the volatile environment requires developers to keep their security knowledge sharp. Even security experts need to get their knowledge updated [36]. Despite the fact that mHealth apps vulnerabilities are frequently announced in the security-relevant knowledge banks (e.g., National Vulnerabilities Database – NVD, data breach reports) to advice developers, for some reasons (i.e., difficult to use), these security alerts are not followed or ignored. As a result, unfixed bugs might allow attackers to perform malicious activities (e.g., illegally access health-critical data by exploiting sensors permissions enabling them to extract data or transfer malware to an app [37]). The announcements of the identified security bugs are one way of encouraging mHealth developers to keep up-to-date with the threat landscape. Muthing et al. and Dehling et al. indicate that mHealth apps developers use out-of-date security measures (S14, S19, S23). As a result, some mHealth apps even have previously exposed security errors (S23). Despite the realization of the importance of keeping mHealth developers aware of the latest security issues, there is a little evidence that developers get regular formal security training to maintain their security knowledge (S24). Lack of auditing among developers to maintain and review their security knowledge can create a knowledge gap, and lead to out-of-date security knowledge (S25).

### C3: Lack of stakeholders' involvement during mHealth app development

Involving stakeholders in security requirement engineering is being recognized as a key to software success and getting effective as well as impactful outcomes [22]. Indeed, stakeholders involvement contributes in the elicitation and specification of security requirements of the developed software, yet it is difficult as developers would first spend a significant effort to understand the complexity of a problem domain [38]. In addition, more time and resources would be required. For mHealth apps, developers should refer to stakeholders (e.g., medics, patients) throughout the development process to ensure that the technology meets their needs (S10, S30). Further, stakeholders need to be involved earlier in the development process of mHealth apps. However, development practices often include clinicians and experts but more rarely involve the target audience until evaluation (S29). At the same time, it would be challenging for some stakeholders to have security understanding due to their capabilities when engaging them in the development process. As a result, causing poor elicitation of security requirements (S5, S20)

## C4: Lack of financial resources for developing secure mHealth app

The development process of mHealth apps can be supported by using security resources to enhance secure mHealth apps development. Lack of necessary resources, such as technology, is a challenge that can directly impact developing secure mHealth apps. For example, security tools (e.g., Zed Attack Proxy, Android Debug Bridge, Codified Security, White Hat Security, and Quick Android Review Kit)[2] are supporting resources to facilitate writing secure code, and testing app during the development process. They help developers catch errors that they might be unaware of and adjust their code accordingly before releasing an app. Wurster et al. [39] argued that not all software developers are security experts, and there is a need to use suitable security tools during a development project. Security tools for mobile apps have received a lot of attention from the researchers. A recent security tool, called FixDroid [40] can show warning messages with recommendations to fix errors during the coding phase. It proved the effectiveness by improving the security of the written code, it is limited to Android apps developers, and it is not widely known.

Similarly, software libraries can be used as supporting resources to facilitate the software development process. Such libraries help developers reuse specific code for certain goals and support access to hardware and software that might be needed. Yet, it can be challenging for developers to know which library to trust while developing mHealth apps. There can be a risk of data leakage by using untrusted libraries (S16, S13). Some libraries, especially the open-source ones, may collect data about users without developers being aware of, leading to data privacy breaches [41]. Furthermore, using untrusted third-party libraries to integrate mHealth app with Electronic Health Record (EHR) can lead attackers to gain unauthorized access to patients' data (S13).

Older versions of security resources (i.e. tools and libraries) also contain known vulnerabilities (S18). Most of the security resources are often updated to address the security-related issues and introduce new functions; hence, it is important to be aware of and use the latest security tools and libraries. Therefore, developers' security knowledge of the adopted security resources can significantly impact the developed app's security. Besides being aware of the relevant security resources, it can be difficult for developers to learn to use them within the time and resources available for a project (S25, S17).

## C5: No/little developers' attention towards the security of mHealth app

Incorporating security should be considered ideally throughout the SDLC from requirement analysis to deployment phase [42]. In fact, addressing security at later stages of app development or after app releasing in the form of security patches can be a costly exercise and can introduce new vulnerabilities [43]. Studies, such as (S11, S21, S31), found that mHealth apps developers pay little or no attention to the security of mHealth app. This issue can be seen for a few reasons including (1) developer' assumption that users are not concerned about security, (2) developers' assumption that security should be handled by app testers, and (3) developers with no security focus would even skip all security measures to resolve other quality attributes including usability and performance (S18, S32). Therefore, it is important to come up with effective mechanisms for overcoming developers' lack of attention towards security.

## C6: Time constraints during mHealth app development process

Due to business pressures (e.g., rushing to the market), delivering an app on time tends to be the main aim mHealth apps developers try to satisfy customers and avoid extra costs. High workload and tight timeframes require mHealth apps developers to put more effort to meet functional requirements as a primary task (S18, S26, S32). It also affects their attitude and behaviour towards addressing security (e.g., underestimating risks, assuming attackers will not realize the weaknesses) and dealing with security after releasing an app [42]. This approach leads to insecure mHealth apps and increases the cost and introduces new vulnerabilities after fixing the existing vulnerabilities [44]. It is estimated that the cost can be 30 to 100 times more expensive to retrofit security compared with incorporating security from the beginning [45]. Besides, the speed of delivering apps will also affect team members to share and convey security knowledge among mHealth apps developers [46]. Furthermore, the long process of gaining consent or approving the developers' choices by their managers can be an issue (S7). As a result, making the process of having their opinion on a certain task take longer. Hence, this lead to skipping security issues that need to be fixed.

---

### C7: Lack of security testing during mHealth app development

Security testing is one of the essential phases of mHealth apps development lifecycle. Security testing helps determine the quality of apps by ensuring all the security requirements are met. Security testing for mHealth apps, in particular, will help to see how an app will react against different attacks (e.g., unauthorized access to health data, tampering health data or reporting invalid health data to health professionals (S11). Security testing of mHealth apps can be overlooked since it can be a challenging task for developers. Several factors can affect performing security testing including the absence of security testing tools, lack of effective and well-known testing guidelines, cost of performing app testing by a third-party organisation, or lack of security expert with a software development organisation (S23, S18, S6). Consequently, this would release mHealth apps without conducting security testing, leaving an app in a high risk [47]. Wurster et al. indicate that security testing is not a first-choice task for developers, and their main job is completing the required features [39].

### C8: Lack of security experts' engagement during mHealth app development

A security expert, security leader or security champion within an organisation plays a vital role during the mHealth apps development process (S7). Besides their development activities, they direct mHealth apps' developers on secure development practices and perform a security review to ensure their code does not have security defects. A security expert can encourage developers to achieve security goals and educate other developers about potential threats and solutions (S14). The lack of security experts within a software development team can lead to failures in applying proper security controls by mHealth apps developers. Besides, the lack of availability of security experts would be a challenge for developers (S7). As a result, lack of constructive feedback that prevents developers from (1) acquiring security knowledge, (2) gaining hands-on experience, and (3) developing apps that are secure by design.

### C9: Developers' lack of motivations and ethical considerations

Motivation refers to the driving force behind all the actions of developers during development. It has been recognized as a critical success factor for software projects. Motivation can be seen differently based on developers and an organisation's size [48]. The research on security practice indicates that many security incidents are mainly caused by human, rather than technical failure [49]. Developers with low motivation were found to be one of the most frequently cited causes of software development project failure [50]. Xie et al. [51] present the reasons that make software developers make security errors. The study concluded that most software developers have "not my problem" attitude, which indicates that software developers are the source of security errors due to their attitudes and behaviours. In mHealth apps development, in particular, studies such as (S10, S27, S30) reported that developers' lack of motivations and ethical consideration is a challenge that hinder developing secure mHealth apps.

## C. Conceptual Framework

Based on our analysis of the extracted data, we propose a conceptual framework, as in Figure 4, that represent the challenges for developing secure mHealth app. Jabareen in [52] defined conceptual framework as "*a network, or a plane of interlinked concepts that together provide a comprehensive understanding of a phenomenon or phenomena.*" Figure 4 presents a conceptual framework for correlating the identified challenges. Next, we present some of the key findings.

### The Role of Security Experts within mHealth Apps Development

*"A critical challenge facing software security today is the dearth of experienced practitioners"— Barnum and McGraw*

A report by IBM showed that there is a dearth of security experts in mobile apps development. Only 41% of the participants indicated that their organizations had sufficient security expertise [45]. Hence, having a security expert can be a strategic advantage for an organization. The role of security experts is quite crucial in developing secure mHealth apps. We conclude from figure 2 that the lack of security experts is already linked with most of the challenges. Without security experts in a team, the required security knowledge will be missing (i.e., what security guidelines need to be followed, what security tools are available to be utilized and which libraries can be trusted). As a result, developers' security knowledge would remain insufficient. Besides, lack of security experts within mHealth apps development organisations can lead to poor coding practices, rushing to deliver an app without even performing security testing. Furthermore, collaboration and social interactions with security experts and other team members would have a significant impact on security. As a result, removing the boundaries and stimulate the formation of common interests, in turn, support exchanging knowledge and ideas [48]. Also, it is a good practice to exchange security knowledge, leverage that knowledge within the project, and acquire new knowledge.

### Importance of Security Knowledge and Expertise for Secure mHealth Apps Development

Our analysis showed that developers' lack of security knowledge and expertise for secure mHealth apps development is correlated with most of the identified challenges. For instance, developing secure mHealth apps requires decent knowledge about security guidelines, security tools, and the trusted libraries (i.e., awareness of how, when and why they should utilize them). It worth to mention that the development of secure mHealth apps has become complex and challenging. mHealth apps require connecting with external sensors or devices, e.g., wearable devices, implantable devices [37]. Nevertheless, providing the required learning resources can be underestimated by mHealth apps development organisations [46]. Thus, organisations are required to provide security material to allow developers to learn to connect mHealth apps with emerging technologies, i.e., Internet of Things (IoT). Providing resources to support secure mHealth app development would contribute to filling the security knowledge gap and help to open developers' mindset to security errors that need to be avoided [36].

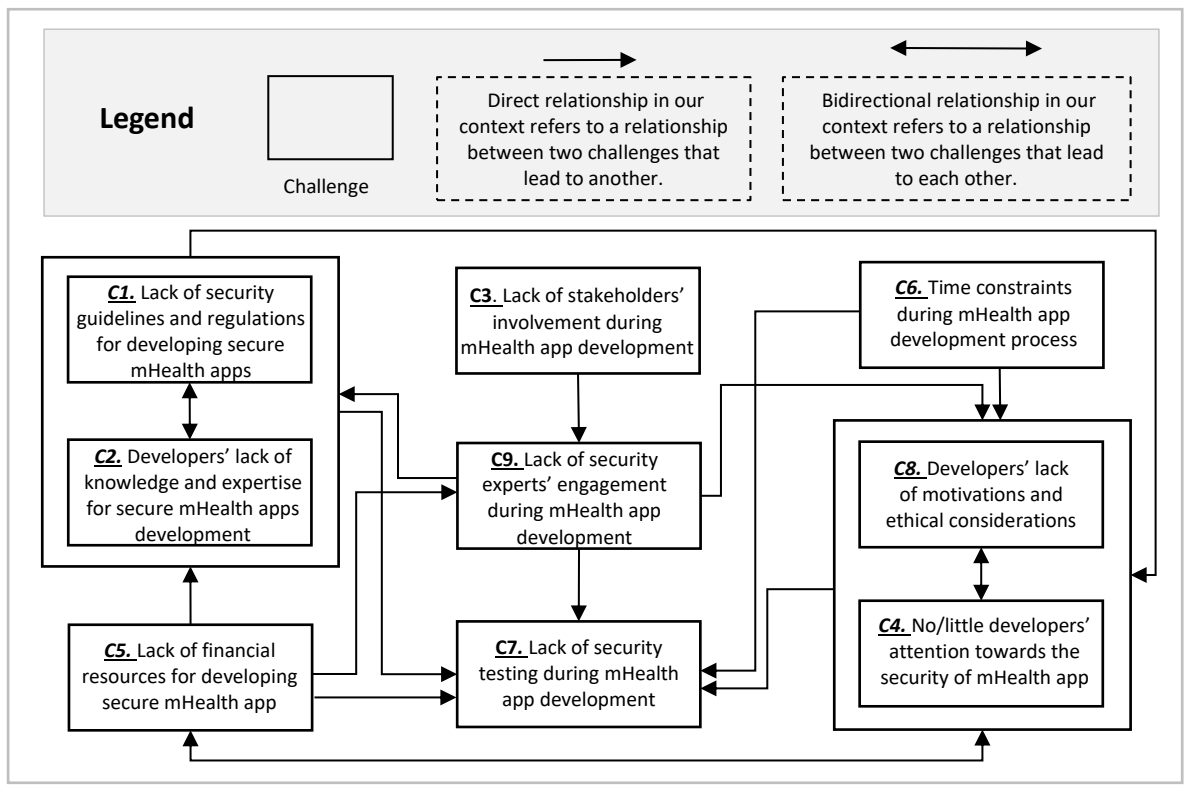

**Figure 4.** A Conceptual Framework for Correlating the Challenges in Developing Secure mHealth Apps

**Frequency Analysis for the Challenges of Developing Secure mHealth Apps**

Critical challenges for developing secure mHealth apps can be determined if a specific challenge has a frequency of ≥50% of the selected studies. This criteria has been used by other researchers in different domains [53, 54]. As in Table 1, the percentages of frequency are shown for each challenge in the reviewed studies. By using this criteria, we conclude that there are two main critical challenges which are: lack of security guidelines for developing secure mHealth apps (20 out of 32 studies, 63%), and developers' lack of security knowledge and expertise for secure mHealth apps development (18 out of 32 studies, 56%). Figure 5 shows the frequency analysis of the identified challenges in the reviewed studies.

Despite the fact that other challenges were given less attention by the reviewed studies (i.e., 19% for C3,

16% for C4, 13% for C5-7, 9% for C8, and 6% for C9); some challenges have a direct relationship with other challenges as we indicated earlier (e.g., poor security decisions during mHealth apps development is related to insufficient security knowledge of developers). Consequently, there will be an impact on the development process of mHealth apps. Therefore, we believe identifying these challenges would help mHealth apps development organizations to evaluate their security practices and readiness in implementing security in mHealth apps projects.

## Discussion and Future Work

We have presented the findings from an SLR in the previous section based upon our research question. In this section, we discuss our findings and highlight some potential future direction.

### Discussion

We noticed from the literature that there is an emphasis on presenting the security issues of mHealth apps, and how they can be resolved (e.g., presenting security framework, providing secure mHealth app development recommendations, evaluating the security for existing mHealth apps, etc.). However, a little attention was given to the human factor during the development process of mHealth apps (i.e., non-technical solutions). Hence, it would be critical to recognize the security challenges that mHealth apps' developers face during the development process.

Sufficient security knowledge for mHealth apps' developers is one of the key factors that would help to to develop secure apps. Security knowledge can be discussed as the type of required security knowledge and the sources of acquiring that knowledge. According to Barnum et al. [33], there are seven security knowledge categories for developing secure software including knowledge of principle, guideline, rule, attack pattern, vulnerability, exploit and historical risk. While the presented set of security knowledge provides a perfect foundation for enhancing security, yet it would be a bit challenging for developers since security knowledge is scattered all around. By considering security knowledge of vulnerabilities as an example, attackers can find a single vulnerability to exploit the app (i.e., launching an attack). In contrast, developers should be aware of all security vulnerabilities and apply proper security measures and patches, which can be a daunting task. mHealth apps, more specifically, are connected to IoT devices which make securing the apps a challenge. Sikder et al. [37] indicated that attackers could illegally access health data through exploiting sensors permissions', which could enable them to extract data and transfer malware to the app. Therefore, further support for mHealth apps' developers' security knowledge is needed to cope with the rapid changes in security knowledge.

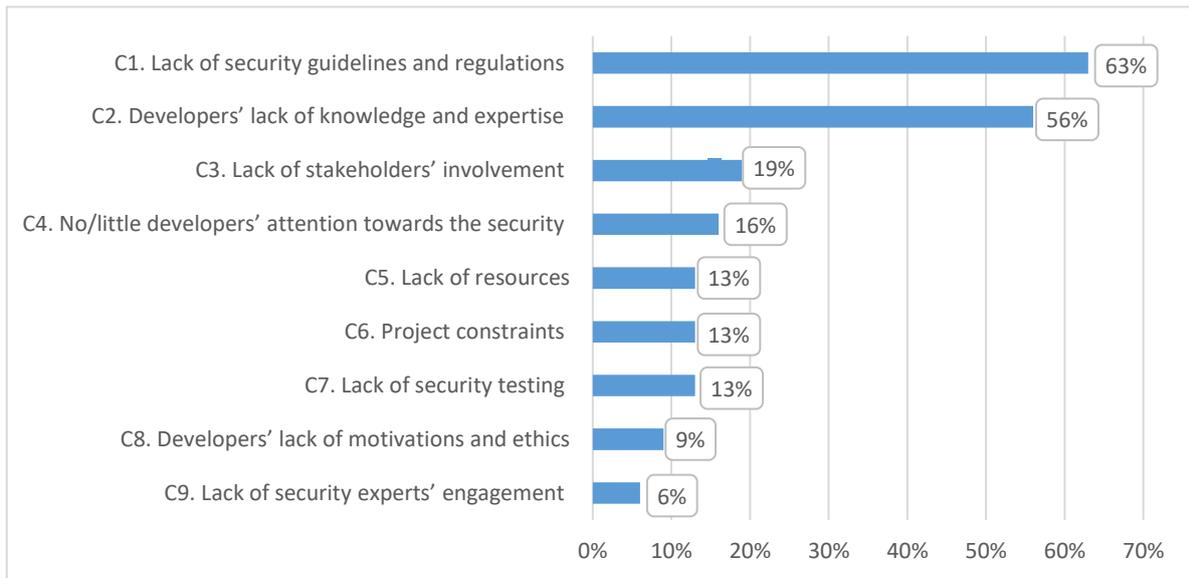

**Figure 5.** Frequency Analysis of the Identified Challenges

> Security knowledge for developers of mHealth apps is an important factor that need to be considered to ensure developing secure apps.

Likewise, using trustable sources (i.e., tools and libraries) would be challenging for developers to have an awareness of their secure usage. So, we suggest further improvement that needs to be done to facilitate the job of mHealth apps' developers by exploring the list of trustable sources. Identifying the trustable sources with their policies, terms and conditions of usage and the proper ways of getting their updates would help mHealth apps' developers to develop secure apps. At the same time, following this approach would help to disseminate and provide security knowledge for mHealth apps developers through trusted sources.

> Necessary resources (e.g., security tools) need to be provided by mHealth apps development organization to help developers deliver secure apps.

## Future Work

The results of our review enabled us to propose the following areas that warrant future research on secure development of mHealth apps

### Challenges and Solutions of Developing Secure mHealth Apps with Real-World Practitioners

In this review, we have identified the challenges that hinder developing secure mHealth apps based on SLR. We plan to conduct an empirical study to investigate the challenges with real-world practitioners to validate our results. The planned future research would enable us to compare the identified challenges identified from the literature with real-world practices for better understanding. Furthermore, we aim to study the practices and solutions that real-world practitioners use to overcome the identified challenges. As a consequence, this would allow us to define which challenges are correlated with which practices. Hence, identifying the challenges and solutions would help us extend the current conceptual framework and provide a body of knowledge for secure mHealth apps development.

### Developers' Motivations and Ethical Considerations for Developing Secure mHealth Apps

Since motivations and ethical considerations play an essential role in secure mHealth apps development process, we assert that there is a need to conduct an empirical study to understand the developers' motivational factors, and what inspires them to ensure the security of mHealth apps (e.g., security leaders, with the team, reward, recognition, career path or promotion). Such a study can be further investigated by collecting quantitative data (e.g., hypothesis testing) and/or qualitative data. This would create a better understanding and help mHealth apps development organizations to realize and focus on motivational factors.

## Study Limitations and Conclusions

### Limitation

One of the potential threats for our SLR can be missing or excluding relevant studies. To mitigate this threat, we used Scopus library as our data source. Scopus considered the largest indexing system that provides the most comprehensive search engine among other digital libraries [55]. Scopus enabled us to get a reasonable number of studies (1867 articles). Furthermore, we tested our search string based on the pilot search to improve it and reach the relevant studies for this review. Study selection was based on a predefined inclusion and exclusion criteria as in Figure 1 (v). However, including and excluding studies can be impacted by the researchers' subjective judgement. To mitigate this threat, the reasons for excluding papers were recorded and reviewed randomly by the second author.

Our research can be influenced by the researcher's bias in extracting data from the reviewed studies, which may negatively affect the findings. To overcome this threat, we extracted data based on a predefined data extraction form (see Appendix 2). Since the first author was done the major part, the second author reviewed and verified the extracted data. Furthermore, data synthesis for the extracted data can impact our results. To mitigate the researchers' bias in interpreting the findings, key points and themes initially done by the first author that was reviewed and revised (followed by a discussion wherever required) by the second author to avoid potential bias.

### Conclusion

This review was motivated by the growing amount of attention to mobile apps, particularly mHealth apps. We aimed to analyze and synthesize the literature to identify the challenges that hinder mHealth apps' developers from developing secure apps. Our review followed an SLR approach and selected 32 studies that we believe are relevant to our study. We have identified and discussed 9 challenges faced by mHealth apps developers to develop secure apps. We also provided a conceptual framework for the identified challenges as well as presented a number of challenges linked to the body of knowledge found in this literature review. Our finding can be valuable for researchers and practitioners (e.g., mHealth app developers, managers) to support research and development of secure mHealth apps. For researchers, this review would help to identify the existing research that would support formulating and testing hypothesis. Furthermore, proposing ideal and innovative solutions to address these challenges. For practitioners, our review would help to understand the existing challenges of developing secure mHealth apps from the literature. This would help to resolve these challenges at the early stages of mHealth apps development process.

## Acknowledgments
We thank Leonardo Horn Iwaya for comments that greatly improved the manuscript. This work is partially funded by Cyber Security Cooperative Research Centre.

## Conflicts of Interest
None declared.

## Multimedia Appendix 1
List of the reviewed studies

| ID | Author(s) | Title | Venue | Pub. Year |
|---|---|---|---|---|
| S1 | S. Gejibo, F. Mancini, K. A. Mughal, R. A. B. Valvik, and J. Klungsøyr | Secure data storage for mobile data collection systems | International Conference on Management of Emergent Digital EcoSystems | 2012 |
| S2 | D. D. Luxton, R. A. Kayl, and M. C. Mishkind | MHealth data security: The need for HIPAA-compliant standardization | Telemedicine and e-Health | 2012 |
| S3 | R. Adhikari, D. Richards, and K. Scott, | Security and privacy issues related to the use of mobile health apps | 25th Australasian Conference on Information Systems | 2014 |
| S4 | D. He, M. Naveed, C. A. Gunter, and K. Nahrstedt | Security Concerns in Android mHealth Apps | Annual Symposium proceedings / AMIA Symposium. | 2014 |
| S5 | T. L. Lewis and J. C. Wyatt | mHealth and mobile medical apps: a framework to assess risk and promote safer use | Journal of medical Internet research | 2014 |
| S6 | S. Becker, T. Miron-Shatz, N. Schumacher, J. Krocza, C. Diamantidis, and U.-V. Albrecht | mHealth 2.0: experiences, possibilities, and perspectives | JMIR mHealth and uHealth | 2014 |
| S7 | I. Mergel | The Long Way From Government Open Data to Mobile Health Apps: Overcoming Institutional Barriers in the US Federal Government | JMIR mHealth and uHealth | 2014 |
| S8 | S. Arora, J. Yttri, and W. Nilsen | Privacy and security in mobile health (mHealth) research | PubMed, Alcohol Research | 2014 |
| S9 | Y. Cifuentes, L. Beltrán, and L. Ramírez | Analysis of Security Vulnerabilities for Mobile Health Applications | The Seventh International Conference on Mobile Computing and Networking | 2015 |
| S10 | A. Carter, J. Liddle, W. Hall, and H. Chenery | Mobile phones in research and treatment: ethical guidelines and future directions | JMIR mHealth and uHealth | 2015 |
| S11 | K. Knorr and D. Aspinall | Security testing for Android mHealth apps | IEEE 8th International Conference on Software Testing, Verification and Validation Workshops, | 2015 |
| S12 | F. Zubaydi, A. Saleh, F. Aloul, and A. Sagahyroon | Security of mobile health (mHealth) systems | IEEE 15th International Conference on Bioinformatics and Bioengineering | 2015 |
| S13 | A. Landman, S. Emani, N. Carlile, I. D. Rosenthal, S. Semakov, J. D. Pallin | A Mobile App for Securely Capturing and Transferring Clinical Images to the Electronic Health Record: Description and Preliminary Usability Study | JMIR mHealth and uHealth | 2015 |
| S14 | T. Dehling, F. Gao, S. Schneider, and A. Sunyaev | Exploring the Far Side of Mobile Health: Information Security and Privacy of Mobile Health Apps on iOS and Android | JMIR mHealth and uHealth | 2015 |
| S15 | J. Hsu, D. Liu, Y. M. Yu, H. T. Zhao, Z. R. Chen, J. Li | The top Chinese mobile health apps: A systematic investigation | Journal of Medical Internet Research | 2016 |
| S16 | D. Kotz, C. A. Gunter, S. Kumar, and J. P. Weiner | Privacy and Security in Mobile Health: A Research Agenda," | Computer, vol. 49, pp. 22-30 | 2016 |
| S17 | M. Zens, N. P. Sï¿½dkamp, and P. Niemeyer | Back on track: cruciate ligament study via smartphone: Practical example of possibilities for the Apple ResearchKit | Arthroskopie, vol. 29 | 2016 |
| S18 | G. Thamilarasu and C. Lakin | A security framework for mobile health applications | 5th International Conference on Future Internet of Things and Cloud Workshops | 2017 |
| S19 | J. Müthing, T. Jäschke, and M. C. Friedrich | Client-Focused Security Assessment of mHealth Apps and Recommended Practices to Prevent or Mitigate Transport Security Issues | JMIR mHealth and uHealth | 2017 |
| S20 | M. Bradway, C. Carrion, B. Vallespin, O. Saadatfard, E. Puigdomènech, M. Espallargues | mHealth Assessment: Conceptualization of a Global Framework | JMIR mHealth and uHealth | 2017 |

| S21 | T. Mabo, B. Swar, and S. Aghili | A vulnerability study of Mhealth chronic disease management (CDM) applications (apps) | Advances in Intelligent Systems and Computing | 2018 |
|---|---|---|---|---|
| S22 | A. Papageorgiou, M. Strigkos, E. Politou, E. Alepis, A. Solanas, and C. Patsakis, | Security and Privacy Analysis of Mobile Health Applications: The Alarming State of Practice | IEEE Access | 2018 |
| S23 | M. Hussain, A. A. Zaidan, B. B. Zidan, S. Iqbal, M. M. Ahmed, O. S. Albahri, et al. | Conceptual framework for the security of mobile health applications on Android platform | Telematics and Informatics | 2018 |
| S24 | L. Hutton, B. A. Price, R. Kelly, C. McCormick, A. K. Bandara, T. Hatzakis, et al. | Assessing the privacy of mhealth apps for self-tracking: heuristic evaluation approach | JMIR mHealth and uHealth | 2018 |
| S25 | M. Aliasgari, M. Black, and N. Yadav | Security vulnerabilities in mobile health applications | IEEE Conference on Application, Information and Network Security | 2019 |
| S26 | Y. M. Al-Sharo | Networking issues for security and privacy in mobile health apps | International Journal of Advanced Computer Science and Applications | 2019 |
| S27 | L. Parker, V. Halter, T. Karliychuk, and Q. Grundy | How private is your mental health app data? An empirical study of mental health app privacy policies and practices | International Journal of Law and Psychiatry | 2019 |
| S28 | M. Srivastava and G. Thamilarasu | MSF: A comprehensive security framework for mhealth applications | International Conference on Future Internet of Things and Cloud Workshops | 2019 |
| S29 | T. Wykes and S. Schueller | Why reviewing apps is not enough: Transparency for trust (T4T) principles of responsible health app marketplaces | Journal of Medical Internet Research | 2019 |
| S30 | T. Wykes, J. Lipshitz, and S. M. Schueller | Towards the Design of Ethical Standards Related to Digital Mental Health and all Its Applications | Current Treatment Options in Psychiatry | 2019 |
| S31 | J. Muchagata, S. Teles, P. Vieira-Marques, D. Abrantes, and A. Ferreira | Dementia and mHealth: On the Way to GDPR Compliance | Communications in Computer and Information Science | 2020 |
| S32 | B. Aljedaani, A. Ahmad, M. Zahedi and M. Ali Babar | An Empirical Study on Developing Secure Mobile Health Apps: The Developers' Perspective | Asia-Pacific Software Engineering Conference | 2020 |

## Multimedia Appendix 2
Data Extraction Form

| # | Data item | Description |
|---|---|---|
| D1 | Author(s) | |
| D2 | Year | Demographic data |
| D3 | The Name of Publication Venue | Demographic data |
| D4 | Title | |
| D5 | Publication Type (i.e., journal, conference, workshop) | Demographic data |
| D6 | Challenges that hinder developing secure mobile health apps. | RQ |

## References

1. Hussain, M., et al., *Conceptual framework for the security of mobile health applications on Android platform.* Telematics and Informatics, 2018.
2. Müthing, J., T. Jäschke, and C.M. Friedrich, *Client-Focused Security Assessment of mHealth Apps and Recommended Practices to Prevent or Mitigate Transport Security Issues.* JMIR mHealth and uHealth, 2017. **5**(10).

3.  Varshney, U., *Mobile health: Four emerging themes of research.* Decision Support Systems, 2014. **66**: p. 20-35.
4.  Aljedaani, B., et al., *An Empirical Study on Developing Secure Mobile Health Apps: The Developers Perspective.* 2020 27th Asia-Pacific Software Engineering Conference (APSEC), 2020.
5.  Knorr, K. and D. Aspinall. *Security testing for Android mHealth apps.* in *2015 IEEE 8th International Conference on Software Testing, Verification and Validation Workshops, ICSTW 2015 - Proceedings.* 2015.
6.  Martínez-Pérez, B., I. de la Torre-Díez, and M. López-Coronado, *Privacy and Security in Mobile Health Apps: A Review and Recommendations.* Journal of Medical Systems, 2015. **39**(1).
7.  Chin, E.M., *Helping developers construct secure mobile applications.* 2013, UC Berkeley.
8.  Research2guidance, *mHealth app Economics: Current Status and Future Trends in Mobile Health retrieved from https://research2guidance.com/product/mhealth-economics-2017-current-status-and-future-trends-in-mobile-health/.* 2017/2018.
9.  Zubaydi, F., et al. *Security of mobile health (mHealth) systems.* in *2015 IEEE 15th International Conference on Bioinformatics and Bioengineering, BIBE 2015.* 2015.
10. Weir, C., B. Hermann, and S. Fahl. *From Needs to Actions to Secure Apps? The Effect of Requirements and Developer Practices on App Security.* in *29th {USENIX} Security Symposium ({USENIX} Security 20).* 2020.
11. Cifuentes, Y., L. Beltrán, and L. Ramírez. *Analysis of Security Vulnerabilities for Mobile Health Applications.* in *2015 Seventh International Conference on Mobile Computing and Networking (ICMCN 2015).* 2015.
12. Adhikari, R., D. Richards, and K. Scott. *Security and privacy issues related to the use of mobile health apps.* in *Proceedings of the 25th Australasian Conference on Information Systems, ACIS 2014.* 2014.
13. Flaten, H.K., et al., *Growth of mobile applications in dermatology-2017 update.* Dermatology online journal, 2018. **24**(2).
14. Zahra, F., A. Hussain, and H. Mohd, *Factor Affecting Mobile Health Application for Chronic Diseases.* Journal of Telecommunication, Electronic and Computer Engineering (JTEC), 2018. **10**(1-11): p. 77-81.
15. Mabo, T., B. Swar, and S. Aghili, *A vulnerability study of Mhealth chronic disease management (CDM) applications (apps),* in *Advances in Intelligent Systems and Computing.* 2018. p. 587-598.
16. Ramey, L., et al., *Apps and Mobile Health Technology in Rehabilitation: The Good, the Bad, and the Unknown.* Physical Medicine and Rehabilitation Clinics, 2019. **30**(2): p. 485-497.
17. Lewis, T.L. and J.C. Wyatt, *mHealth and mobile medical apps: a framework to assess risk and promote safer use.* Journal of medical Internet research, 2014. **16**(9).
18. Dehling, T., et al., *Exploring the far side of mobile health: information security and privacy of mobile health apps on iOS and Android.* JMIR mHealth and uHealth, 2015. **3**(1).
19. Kotz, D., et al., *Privacy and Security in Mobile Health: A Research Agenda.* Computer, 2016. **49**(6): p. 22-30.
20. Papageorgiou, A., et al., *Security and Privacy Analysis of Mobile Health Applications: The Alarming State of Practice.* IEEE Access, 2018. **6**: p. 9390-9403.
21. Gejibo, S., et al. *Secure data storage for mobile data collection systems.* in *Proceedings of the International Conference on Management of Emergent Digital EcoSystems, MEDES 2012.* 2012.
22. Kanniah, S.L. and M.N.r. Mahrin, *A Review on Factors Influencing Implementation of Secure Software Development Practices.* World Academy of Science, Engineering and Technology, International Journal of Social, Behavioral, Educational, Economic, Business and Industrial Engineering, 2016. **10**(8): p. 3022-3029.
23. Thomas, T.W., et al. *Security During Application Development: an Application Security Expert Perspective.* in *Proceedings of the 2018 CHI Conference on Human Factors in Computing Systems.* 2018. ACM.
24. Raghavan, V. and X. Zhang, *Building security in during information systems development.* AMCIS 2009 Proceedings, 2009: p. 687.
25. Weir, C., A. Rashid, and J. Noble, *Developer Essentials: Top Five Interventions to Support Secure Software Development.* 2017, Lancaster University.
26. Acar, Y., et al. *Developers Need Support, Too: A Survey of Security Advice for Software Developers.* in *Cybersecurity Development (SecDev), 2017 IEEE.* 2017. IEEE.
27. Chatzipavlou, I.A., S.A. Christoforidou, and M. Vlachopoulou, *A recommended guideline for the development of mHealth Apps.* Mhealth, 2016. **2**.
28. Katusiime, J. and N. Pinkwart, *A review of privacy and usability issues in mobile health systems: Role of external factors.* Health Informatics Journal, 2019. **25**(3): p. 935-950.
29. Marquez, G., H. Astudillo, and C. Taramasco, *Security in Telehealth Systems from a Software Engineering Viewpoint: A Systematic Mapping Study.* IEEE Access, 2020. **8**: p. 10933-10950.
30. Kitchenham, B., et al., *Systematic literature reviews in software engineering–a tertiary study.* Information and Software Technology, 2010. **52**(8): p. 792-805.
31. Cruzes, D.S. and T. Dyba. *Recommended steps for thematic synthesis in software engineering.* in *Empirical Software Engineering and Measurement (ESEM), 2011 International Symposium on.* 2011. IEEE.

32. Regoniel, P.A., *Conceptual framework: A step by step guide on how to make one.* SimplyEducate. Me, 2015.
33. Barnum, S. and G. McGraw, *Knowledge for software security.* IEEE Security & Privacy, 2005. **3**(2): p. 74-78.
34. Nurgalieva, L., D. O'Callaghan, and G. Doherty, *Security and Privacy of mHealth Applications: A Scoping Review.* IEEE Access, 2020. **8**: p. 104247-104268.
35. Assal, H. and S. Chiasson. *'Think secure from the beginning' A Survey with Software Developers.* in *Proceedings of the 2019 CHI Conference on Human Factors in Computing Systems.* 2019.
36. Weir, C., A. Rashid, and J. Noble, *How to Improve the Security Skills of Mobile App Developers: Comparing and Contrasting Expert Views.* 2016.
37. Sikder, A.K., et al., *A Survey on Sensor-based Threats to Internet-of-Things (IoT) Devices and Applications.* arXiv preprint arXiv:1802.02041, 2018.
38. Faily, S., *Engaging Stakeholders in Security Design: An Assumption-Driven Approach.* 2014.
39. Wurster, G. and P.C. van Oorschot. *The developer is the enemy.* in *Proceedings of the 2008 New Security Paradigms Workshop.* 2009. ACM.
40. Nguyen, D.C., et al., *A Stitch in Time: Supporting Android Developers in Writing Secure Code.* 2017.
41. Balebako, R., et al., *The privacy and security behaviors of smartphone app developers.* 2014.
42. McGraw, G., *Software security: building security in.* Vol. 1. 2006: Addison-Wesley Professional.
43. Howard, M. and S. Lipner, *The security development lifecycle.* Vol. 8. 2006: Microsoft Press Redmond.
44. Woon, I.M. and A. Kankanhalli, *Investigation of IS professionals' intention to practise secure development of applications.* International Journal of Human-Computer Studies, 2007. **65**(1): p. 29-41.
45. IBM, *The State of Mobile Application Insecurity available at https://www-01.ibm.com/common/ssi/cgi-bin/ssialias?htmlfid=WGL03074BEEN.* 2015.
46. centre, N.c.s., *Secure development is everyone's concern available at https://www.ncsc.gov.uk/guidance/secure-development-everyones-concern.* 2017.
47. Inukollu, V.N., et al., *Factors influencing quality of mobile apps: Role of mobile app development life cycle.* arXiv preprint arXiv:1410.4537, 2014.
48. Sharp, H., et al., *Models of motivation in software engineering.* Information and software technology, 2009. **51**(1): p. 219-233.
49. Beautement, A., M.A. Sasse, and M. Wonham. *The compliance budget: managing security behaviour in organisations.* in *Proceedings of the 2008 New Security Paradigms Workshop.* 2009. ACM.
50. Verner, J.M., et al., *Factors that motivate software engineering teams: A four country empirical study.* Journal of Systems and Software, 2014. **92**: p. 115-127.
51. Xie, J., H.R. Lipford, and B. Chu. *Why do programmers make security errors?* in *Visual Languages and Human-Centric Computing (VL/HCC), 2011 IEEE Symposium on.* 2011. IEEE.
52. Jabareen, Y., *Building a conceptual framework: philosophy, definitions, and procedure.* International journal of qualitative methods, 2009. **8**(4): p. 49-62.
53. Khan, A.A., et al., *Understanding software process improvement in global software development: a theoretical framework of human factors.* ACM SIGAPP Applied Computing Review, 2017. **17**(2): p. 5-15.
54. Niazi, M., D. Wilson, and D. Zowghi, *Critical success factors for software process improvement implementation: an empirical study.* Software Process: Improvement and Practice, 2006. **11**(2): p. 193-211.
55. Shahin, M., M.A. Babar, and L. Zhu, *Continuous integration, delivery and deployment: a systematic review on approaches, tools, challenges and practices.* IEEE Access, 2017. **5**: p. 3909-3943.

**Abbreviations**
**mHealth:** mobile health
**Apps:** Applications
**HIPAA:** Health Insurance Portability and Accountability Act
**GDPR:** European General Data Protection Regulation
**WHO:** World Health Organization
**SDLC:** Software Development Lifecycle
**SLR:** Systematic Literature Review
**EBSE:** Evidence-Based Software Engineering
**EHR:** Electronic Health Record
**IoT:** Internet of Things
**API:** Application Programming Interface
**WSNs:** Wireless Sensor Networks
**NVD:** National Vulnerabilities Database
**OWASP:** Open Web Application Security Project